\documentclass[twocolumn]{aastex631}
\usepackage{amsmath}

\received{September 4, 2022}
\revised{November 30, 2022}
\accepted{December 5, 2022}

\publishjournal{Astrophysical Journal Letters \\ DOI: \url{https://doi.org/10.3847/2041-8213/aca90c}}

\begin{document}

\title{Diverse Carbonates in Exoplanet Oceans Promote the Carbon Cycle}

\correspondingauthor{Kaustubh Hakim}
\email{kaustubh.hakim@unibe.ch}

\author[0000-0003-4815-2874]{Kaustubh Hakim}
\affiliation{University of Bern, Center for Space and Habitability, Gesellschaftsstrasse 6, CH-3012 Bern, Switzerland}

\author[0000-0002-7384-8577]{Meng Tian}
\affiliation{University of Bern, Center for Space and Habitability, Gesellschaftsstrasse 6, CH-3012 Bern, Switzerland}

\author[0000-0002-0673-4860]{Dan J. Bower}
\affiliation{University of Bern, Center for Space and Habitability, Gesellschaftsstrasse 6, CH-3012 Bern, Switzerland}

\author[0000-0003-1907-5910]{Kevin Heng}
\affiliation{Ludwig Maximilian University, University Observatory Munich, Scheinerstrasse 1, Munich D-81679, Germany}
\affiliation{University of Warwick, Department of Physics, Astronomy \& Astrophysics Group, Coventry CV4 7AL, United Kingdom}
\affiliation{University of Bern, ARTORG Center for Biomedical Engineering Research, Murtenstrasse 50, CH-3008, Bern, Switzerland}



\begin{abstract}

Carbonate precipitation in oceans is essential for the carbonate-silicate cycle (inorganic carbon cycle) to maintain temperate climates. By considering the thermodynamics of carbonate chemistry, we demonstrate that the ocean pH decreases by approximately 0.5 for a factor of 10 increase in the atmospheric carbon dioxide content. The upper and lower limits of ocean pH are within 1--4 of each other, where the upper limit is buffered by carbonate precipitation and defines the ocean pH when the carbon cycle operates. If the carbonate compensation depth (CCD) resides above the ocean floor, then carbonate precipitation and the carbon cycle cease to operate. The CCD is deep ($>$40 km) for high ocean temperature and high atmospheric carbon dioxide content. Key divalent carbonates of magnesium, calcium and iron produce an increasingly wider parameter space of deep CCDs, suggesting that chemical diversity promotes the carbon cycle. The search for life from exoplanets will benefit by including chemically more diverse targets than Earth twins. 

\end{abstract}

\keywords{Extrasolar rocky planets (511); Carbon dioxide (196); Habitable zone (696); Ocean-atmosphere interactions (1150); Geological processes (2288); (\textit{\small Unified Astronomy Thesaurus}) }


\section{Introduction} \label{sec:intro}

The carbonate-silicate cycle, also known as the inorganic carbon cycle, is a negative climate feedback mechanism that stabilises 
the surface temperature via the greenhouse effect of carbon dioxide in response to changes in volcanism rates, stellar luminosity, atmospheric composition and opacity, planetary orbital movements and spin axis tilt \citep{2004tpcc.book.....B,2017aeil.book.....C}. Continental silicate rocks and atmospheric carbon dioxide react with water in a process known as silicate weathering to produce carbonate-forming ions that precipitate as carbonates onto the ocean floor \citep{1981JGR....86.9776W}. The carbon cycle is completed when carbonates are transferred into the mantle for deep storage or carbon is eventually released back into the atmosphere by volcanism \citep{1978CAO..book.....B,2001JGR...106.1373S}, although the degassing efficiency is debated \citep{2015PNAS..112E3997K,2015ApJ...812...36F}. Silicate weathering and carbonate precipitation are traditionally represented by the net chemical reaction \citep{1981JGR....86.9776W},
\begin{equation}
\mbox{CaSiO}_3 + \mbox{CO}_2 \rightarrow \mbox{CaCO}_3 + \mbox{SiO}_2,
\end{equation}
where wollastonite (CaSiO$_3$), which serves as a proxy for silicate rocks, is converted into calcite (CaCO$_3$). Calcium thus plays a crucial role in silicate weathering and carbonate precipitation and is present as Ca$^{2+}$ cations in oceans (Sect. \ref{sec:methods}).

\begin{figure}
\begin{center}
\includegraphics[width=0.49\textwidth]{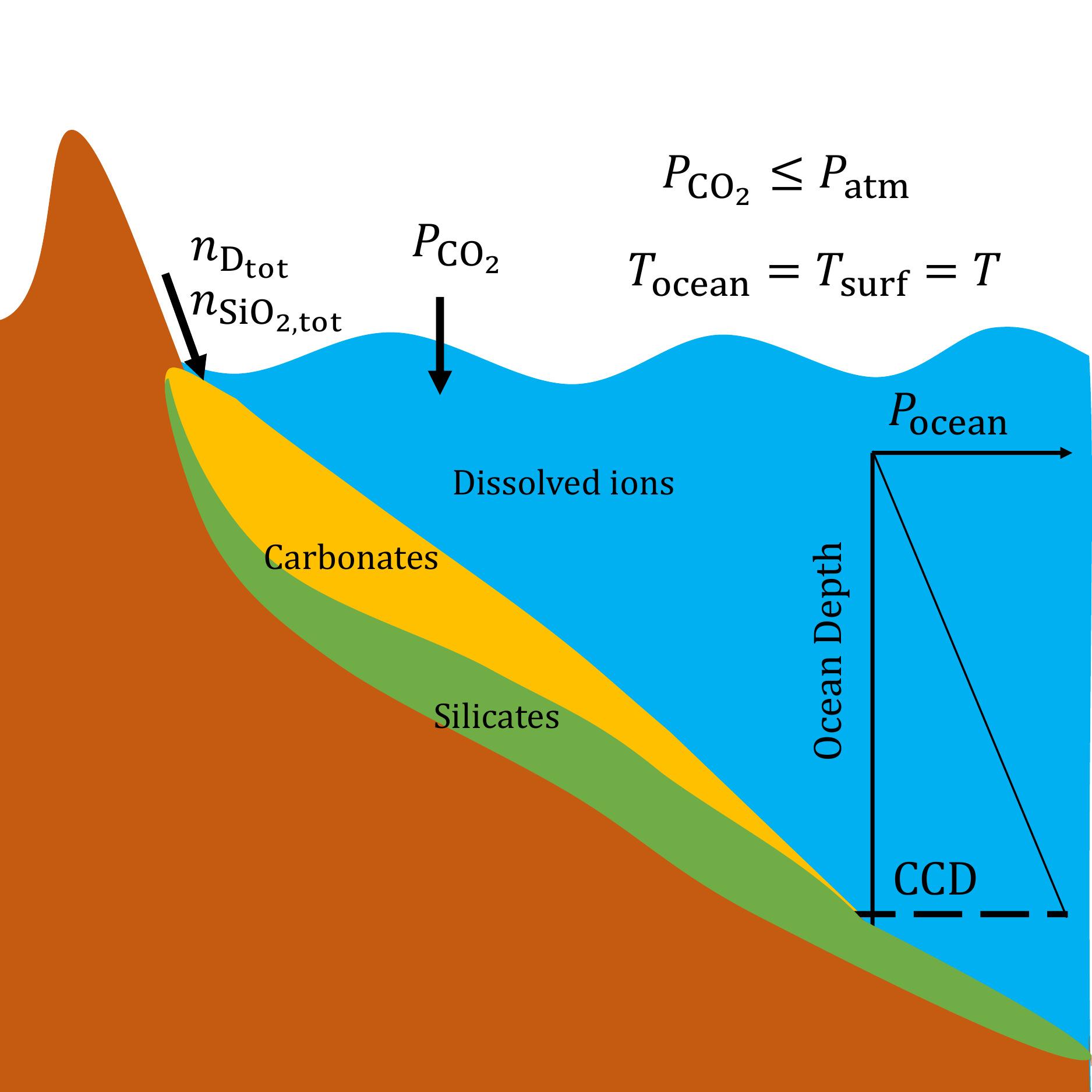}
\end{center}
\vspace{-0.2in}
\caption{Model parameters, $n_{\rm D_{tot}}$ where D = Ca, Mg or Fe, $n_{\rm SiO_{2,tot}}$, $P_{\rm CO_2}$, $P_{\rm atm}$, $P_{\rm oc}$ and $T$. See Sect. \ref{sec:methods} and Table~\ref{tab:params} for a full list of output quantities and description.}
\vspace{-0.1in}
\label{fig:fig1}
\end{figure}

The existence of habitable zones assumes that the carbon cycle operates on Earth analogues to stabilise their atmospheric carbon dioxide content \citep{1993Icar..101..108K}.  Implicitly, this assumes not only that silicate weathering operates, but that ocean floor precipitation and deep storage of carbonates also occur. There exists a critical ocean depth known as the carbonate compensation depth (CCD), below which carbonates are unable to exist in their solid form because carbonate solubility increases with pressure in the ocean \citep[][see also Sect.  \ref{sec:methodsCCD}, Figure~\ref{fig:fig1}]{2003GGG.....4.1104Z}. In modern Earth oceans, the CCD is located between 4--5~km, below the average ocean depth of about 3.8~km \citep{2012AREPS..40..141Z}. If the CCD resides at a depth above the ocean floor, then carbonates are unable to settle.  This leads to the disruption of the carbon cycle---at least, as it is understood to operate on Earth. Moreover, there are currently no theoretical constraints on exoplanet ocean chemistry. We investigate the interplay between atmospheric carbon dioxide content, ocean acidity (pH) and carbonate precipitation. We then calculate the CCD over a broad range of physical conditions. 

\section{Methods} \label{sec:methods}

\subsection{Ocean chemistry model} \label{sec:methodsOceanChem}

\subsubsection{Ca system} 

Ocean chemistry is modelled by considering thermochemical equilibrium for pure Ca, Mg, or Fe systems. The CO$_2$ partial pressure $P_{\rm CO_2}$, ocean--surface temperature $T$ and local ocean pressure $P_{\rm oc}$ are control parameters (Figure~\ref{fig:fig1}, Table~\ref{tab:params}). In the Ca system, there are 13 unknowns, the number density $n$ of H$^+$, OH$^-$, H$_2$O, HCO$_3^-$, CO$_3^{2-}$, CO$_2(aq)$,  Ca$_{\rm tot}$, Ca$^{2+}$, SiO$_{\rm 2,tot}$, SiO$_2(aq)$, quartz SiO$_2(s)$, wollastonite CaSiO$_3(s)$ and calcite CaCO$_3(s)$. Out of the 13 unknowns, 2 are continental silicate weathering products, $n_{\rm Ca_{tot}}$ and $n_{\rm SiO_{2, tot}}$, that depend on $P_{\rm CO_2}$ and $T$ (Sect. \ref{sec:methodsWeathModel}). There are 11 remaining unknowns. We solve for 3 mass conservation equations (for H, Ca and SiO$_2$), 1 charge balance equation, and 7 equations from 7 chemical reactions providing relations between equilibrium constants (that depend on $P_{\rm oc}$ and $T$), reactants and products. 
\begin{center}
\begin{table}[ht!]
\footnotesize
\caption{Parameters and output quantities. \label{tab:params}} 
\begin{tabular}{llc}
\hline
Symbol & Description & Reference \\
\hline
& \textit{Parameters for ocean chemistry}\\
$T$    & Ocean--Surface temperature &  288 K      \\
$P_{\rm{CO_2}}$ & CO$_2$ partial pressure & 0.3 mbar \\
$P_{\rm atm}$    & Atmospheric pressure & 1 bar    \\
$P_{\rm oc}$  & Ocean layer pressure & 1 bar    \\

\hline
& \textit{Parameters for weathering }\\
$n_{\rm{D_{tot,0}}}$ & Ca, Mg or Fe ref. number density  & 1 m$^{-3}$ \\
$\beta$ & Weathering power-law exponent & 0.3 \\
$T_e$ & $e$-folding temperature & 13.7~K \\

\hline
& \textit{Output quantities}\\
$n_X$  & Number density of $X$ [m$^{-3}$] \\
pH     & --log$_{10}$($n_{\rm H^+}/n_0$); $n_0 = 10^3$~m$^{-3}$  \\
\hline
\end{tabular}
\end{table}
\end{center}
These 3 mass-conservation equations, 1 charge balance equation and 7 reactions (water dissociation, Henry's law/physical CO$_2$ dissolution, chemical CO$_2$ dissolution, bicarbonate ion dissociation, calcite precipitation, quartz precipitation and wollastonite precipitation) are specified below. Henry's law gives the amount of CO$_2$ physically dissolved in ocean water in equilibrium with $P_{\rm CO_2}$:
\begin{equation}
\label{eq:HenryCO2}
\mathrm{CO}_2(g) \rightleftharpoons \mathrm{CO}_2(aq).
\end{equation}
The chemical dissolution or dissociation of CO$_2$ in ocean water leads to the production of HCO$_3^-$ and H$^+$ ions and thereby increases the ocean acidity (and decreases ocean pH = $- \log_{10} (n_{\rm H^+} / n_0$), where the standard number density $n_0 = 1$~m$^{-3}$) by the following reaction:
\begin{equation}
\label{eq:ChemicalCO2}
\mathrm{CO}_2(g) + \rm{H_2O} \rightleftharpoons \rm{H^+} + \rm{HCO_3^-}.
\end{equation}
To maintain the charge balance in ocean water, the addition of Ca$^{2+}$ to oceans decreases the number density of H$^+$ and hence increases the ocean pH. The charge balance equation is given by:
\begin{equation}
\label{eq:ChargeBalance}
2 n_{\rm Ca^{2+}} + n_{\rm H^{+}} =  n_{\rm HCO_3^-} + 2 n_{\rm CO_3^{2-}} + n_{\rm OH^-},
\end{equation}
where ${\rm CO_3^{2-}}$ is produced due to the bicarbonate dissociation reaction:
\begin{equation}
\label{eq:BicarbDiss}
\rm HCO_3^- \rightleftharpoons \rm CO_3^{2-} + H^+,
\end{equation}
and where ${\rm OH^-}$ is produced due to the water dissociation reaction:
\begin{equation}
\label{eq:WaterDiss}
\rm H_2O \rightleftharpoons \rm H^+ + OH^-.
\end{equation}
The mass conservation of H is given by 
\begin{equation}
\label{eq:HMass}
n_{\rm H_{tot}} = 2 n_{\rm H_2O} + n_{\rm H^{+}} +  n_{\rm HCO_3^-}.
\end{equation}
Ca$_{\rm tot}$ partitions into Ca$^{2+}$, calcite and wollastonite which is accounted for by mass conservation:
\begin{equation}
\label{eq:CaMass}
n_{\rm Ca_{tot}} = n_{\rm Ca^{2+}} + n_{\rm Cal} + n_{\rm Wo}.
\end{equation}
Calcite precipitation occurs when $n_{\rm Ca^{2+}}$ is saturated to a certain value determined by the equilibrium constant of the calcite precipitation reaction and the abundance of $n_{\rm CO_3^{2-}}$:
\begin{equation}
\label{eq:Ca++}
\rm Ca^{2+} + CO_3^{2-} \rightleftharpoons \mathrm{CaCO}_3(s).
\end{equation}
SiO$_{\rm 2, tot}$ partitions into aqueous silica SiO$_2(aq)$, quartz SiO$_2(s)$ and wollastonite CaSiO$_3(s)$. The mass conservation for SiO$_2$ is given by:
\begin{equation}
\label{eq:SiO2Mass}
n_{\rm SiO_{2,tot}} = n_{\mathrm{SiO}_2(aq)} + n_{\mathrm{Qz}} + n_{\rm Wo}.
\end{equation}
The quartz precipitation reaction is: 
\begin{equation}
\label{eq:SiO2s}
\mathrm{SiO}_2(aq) \rightleftharpoons \mathrm{SiO}_2(s).
\end{equation}
The reaction of wollastonite precipitation is given by: 
\begin{equation}
\label{eq:CaSiO3}
\mathrm{Ca}^{2+} + \mathrm{SiO}_2(aq) + \rm H_2O \rightleftharpoons \mathrm{2 H}^{+} + \mathrm{CaSiO}_3(s).
\end{equation}

These equilibrium chemistry calculations are performed using \texttt{Reaktoro v2} \citep{2015reaktoro}, a multi-phase (aqueous, gas and solid mineral phases) chemistry software. This software implements the extended law of mass action including the determination of stable and unstable species for a given set of species in the system \citep{2017Leal}. We use the SUPCRTBL database for thermodynamic data \citep{1992CG.....18..899J,2016CG.....90...97Z}, the Peng-Robinson activity model for gases \citep{1976PengRobinson}, the HKF activity model for water \citep{1981AmJS..281.1249H} and the Drummond activity model for CO$_2(aq)$ \citep{1981Drummond}. \\

\subsubsection{Mg and Fe systems}

In the Mg system, Ca is replaced by Mg, calcite by magnesite MgCO$_3(s)$ and wollastonite by enstatite Mg$_2$Si$_2$O$_6(s)$. This includes replacing equilibrium constants of all reactions including Mg. Similarly, in the Fe system, Ca is replaced by Fe, calcite by siderite FeCO$_3(s)$ and wollastonite by fayalite Fe$_2$SiO$_4(s)$.
We limit our calculations to Fe$^{2+}$ although its oxidation has inhibited the formation of siderite during Earth's history, particularly since the great oxidation event \citep{1995Natur.378..603R}.

\subsection{Weathering model}
\label{sec:methodsWeathModel}

The introduction of carbonate-producing divalent cations in oceans is dictated by silicate weathering. Silicate weathering and therefore the total number density of divalent cations D$^{2+}$ (D = Ca, Mg or Fe) must depend on the CO$_2$ partial pressure $P_{\rm CO_2}$ and surface temperature $T$ \citep{1981JGR....86.9776W,2021PSJ.....2...49H},

\begin{equation} \label{eq:funcCa}
\small
    n_{\rm D_{tot}} = f_{W} {\scriptsize (P_{\rm CO_2}, T)} = n_{\rm D_{tot,0}} {\scriptsize \left( \frac{P_{\rm CO_2}}{P_{\rm CO_2, 0}} \right)^{\beta} \exp{\left( \frac{T - T_0}{T_e} \right)} },
\end{equation}
where `0' represents the Earth reference values (Table~\ref{tab:params}), $T_e = 13.7$~K is the $e$-folding temperature and $\beta = 0.3$ is the weathering power-law exponent \citep{1981JGR....86.9776W}. However, not all added Ca (or Mg, Fe) in oceans remains in the form of divalent cations, a fraction of it precipitates as carbonates on the ocean floor and another fraction as silicates. For this reason, we perform partitioning calculations of Ca (or Mg, Fe) in different phases following the ocean chemistry model (Sect. \ref{sec:methodsOceanChem}).

\subsection{CCD model} \label{sec:methodsCCD}

Carbonates are deposited onto the ocean floor as part of sediments. The transition from calcite-rich to calcite-free sediments is gradual. The carbonate compensation depth (CCD) for the Earth ocean is normally defined as the depth at which 
the dissolution flux of calcite balances the precipitation flux \citep{2012AREPS..40..141Z}. The depth at which the rapid dissolution of calcite-rich sediments begins is known as the lysocline, which is a sediment property \citep{2003GGG.....4.1104Z}. The lysocline and CCD serve as bounds on the transition zone ($\sim$0.5~km) between calcite-rich and calcite-free sediments. Other definitions for the CCD exist \citep{1976JGR....81.2617B,2005E&PSL.234..299R,2012AREPS..40..141Z}. The depth of ocean $d$ [km] in terms of ocean pressure $P_{\rm oc}$ [bar] at the equator is given by \citep{1998ASAJ..103.1346L}

\begin{equation} \label{eq:ocdepth}
\small
\begin{aligned}
    d = &\frac{1}{9.7803 \times 10^3 + 0.011 P_{\rm oc}} 
    (97.266 P_{\rm oc} - 2.512 \times 10^{-3} P_{\rm oc}^2 \\
    & + 2.28 \times 10^{-7} P_{\rm oc}^3 - 1.8 \times 10^{-11} P_{\rm oc}^4 ).
\end{aligned}
\end{equation}
We consider the CCD to be the depth $d_{\rm CCD}$ (equivalent to the ocean pressure where $P_{\rm oc} = P_{\rm CCD}$) at which 99.9\% of near-surface ($P_{\rm oc} = P_{\rm surf}$) Ca, Mg or Fe-carbonates dissolve,

\begin{equation} \label{eq:CCD}
    n_{\rm Carb, CCD} = 0.001~n_{\rm Carb, surf}.
\end{equation}
Our calculations of CCD are performed up to $d_{\rm CCD} = 45$~km because of the availability of thermodynamic data up to the pressure of 5000~bar \citep{2016CG.....90...97Z}. This limitation does not affect our conclusions.

\subsection{Analytical solution of ocean pH} \label{sec:methodsAnalytic}

\textit{Upper limit of ocean pH.} For calcite precipitation, all reactions in Section \ref{sec:methods} need to be satisfied. However, two of these reactions can be used to analytically constrain ocean pH: Equations \ref{eq:Ca++} and \ref{eq:ChemicalCO2B} where Equation \ref{eq:ChemicalCO2B} is a combination of Equations \ref{eq:ChemicalCO2} and \ref{eq:BicarbDiss},

\begin{equation}
\label{eq:ChemicalCO2B}
\mathrm{CO}_2(g) + \rm{H_2O} \rightleftharpoons \rm{2H^+} + \rm{CO_3^{2-}}.
\end{equation}
The ocean pH can be written as a function of $P_{\rm CO_2}$, $n_{\rm Ca^{2+}}$ and equilibrium constants of Equations \ref{eq:Ca++} and \ref{eq:ChemicalCO2B} (Appendix \ref{app:A}):

\begin{equation} \label{eq:pHUpper}
    \mathrm{pH} = -\frac{1}{2} \left( \log P_{\rm CO_2} + \log K_9 K_{16} + \log \frac{n_{\rm Ca^{2+}}}{n_0} \right).
\end{equation}
This equation demonstrates the reason for the slope of approximately $-0.5$ for the upper limit of ocean pH as a function of the logarithm (base 10) of $P_{\rm CO_2}$. Because $K_9$ and $K_{16}$ are constants at a fixed $T$ and $P$, pH becomes a function of only $P_{\rm CO_2}$ and $n_{\rm Ca^{2+}}$ in Equation \ref{eq:pHUpper}. As a function of $P_{\rm CO_2}$, $n_{\rm Ca^{2+}}$ at the limit of carbonate saturation varies between $\sim$0.1 m$^{-3}$ (at $P_{\rm CO_2} = 0.01$~$\mu$bar) and $\sim$6 m$^{-3}$ (at $P_{\rm CO_2} = 0.3$~bar). This additional increase in $n_{\rm Ca^{2+}}$ of less than two orders of magnitude over seven orders of magnitude increase in $P_{\rm CO_2}$, makes the slope of ocean pH slightly steeper than $-0.5$ (see Fig.~\ref{fig:figA1}). Using $n_{\rm Ca^{2+}}$ from the numerical solution in Equation \ref{eq:pHUpper} results in a semi-analytical solution matching with the numerical solution until $P_{\rm CO_2} = 0.1$~bar, beyond which non-ideal effects accounted in the numerical solution exhibit a small deviation from the analytical equation. 

\textit{Lower limit of ocean pH.} In the absence of divalent cations in ocean, the ocean pH is largely governed by the conversion of CO$_2$ to protons (Equation \ref{eq:ChemicalCO2}). For $P_{\rm CO_2} > 1$~$\mu$bar, the ocean is acidic, where the number density of H$^+$ is larger than that of OH$^-$ and the number density of HCO$_3^-$ is larger than CO$_3^{2-}$  \citep[bicarbonate-carbonate-water equilibria,][]{2007Wolf}. Therefore, the charge balance equation can be approximated as 
\begin{equation}
\label{eq:ChargeBalanceC}
n_{\rm H^{+}} = n_{\rm HCO_3^-}.
\end{equation}
In terms of the equilibrium constant of Equation \ref{eq:ChemicalCO2}, this leads to  (Appendix \ref{app:A})
\begin{equation} \label{eq:pHLower}
    \mathrm{pH} = -\frac{1}{2} \left( \log P_{\rm CO_2} + \log K_3 \right).
\end{equation}
At a fixed $T$ and $P$, $K_3$ is constant and thus the ocean pH exhibits a slope of $-0.5$ for $P_{\rm CO_2} > 1$~$\mu$bar (Fig.~\ref{fig:figA1}). For $P_{\rm CO_2} < 1$~$\mu$bar, the analytical solution does not hold because the number density of OH$^-$ is significant enough to make the charge balance approximation in Equation~\ref{eq:ChargeBalanceC} invalid. The lower limit of ocean pH is independent of the Ca, Mg or Fe systems considered. 

\begin{figure}
\begin{center}
\includegraphics[width=0.375\textwidth]{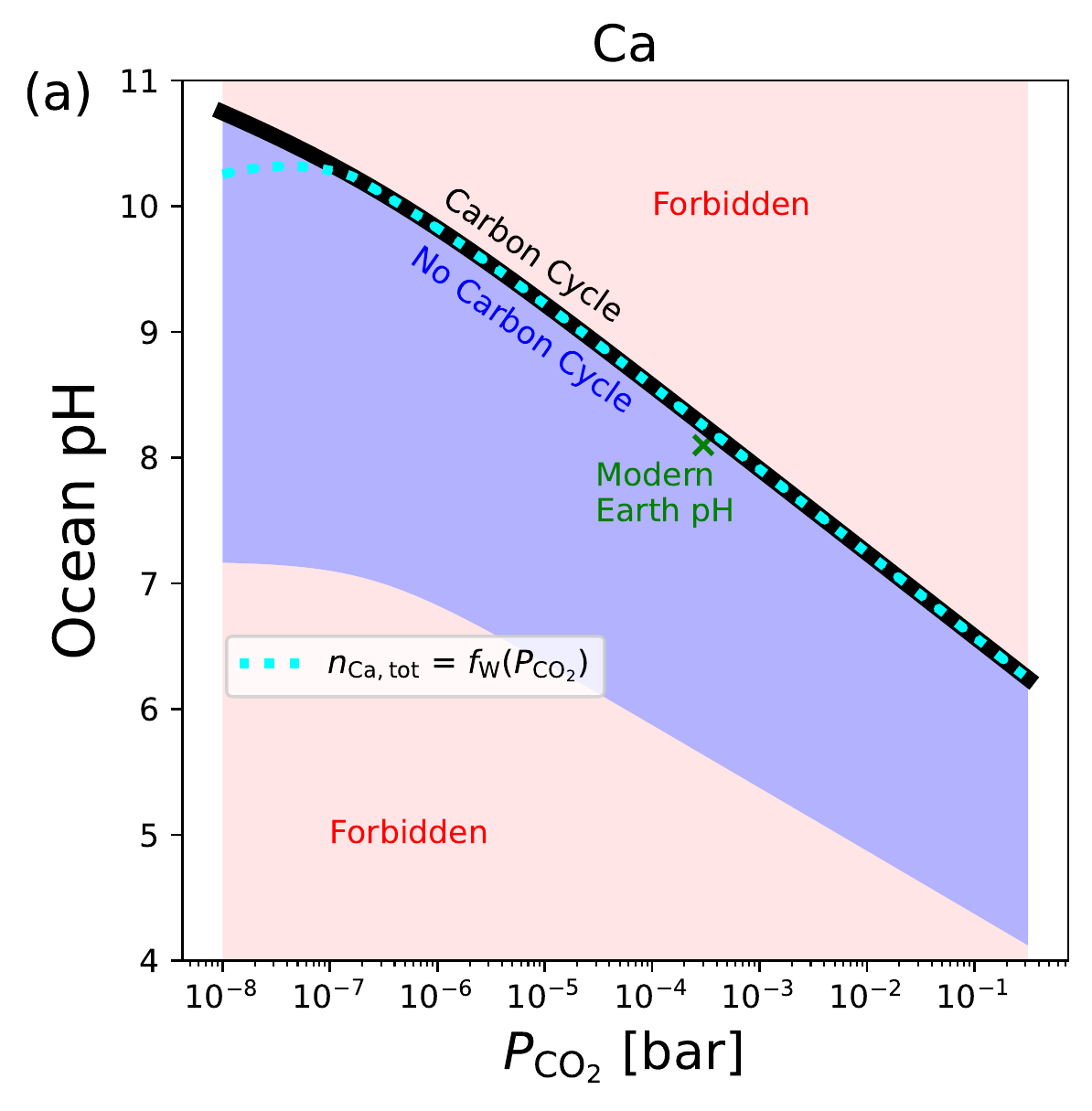}
\includegraphics[width=0.375\textwidth]{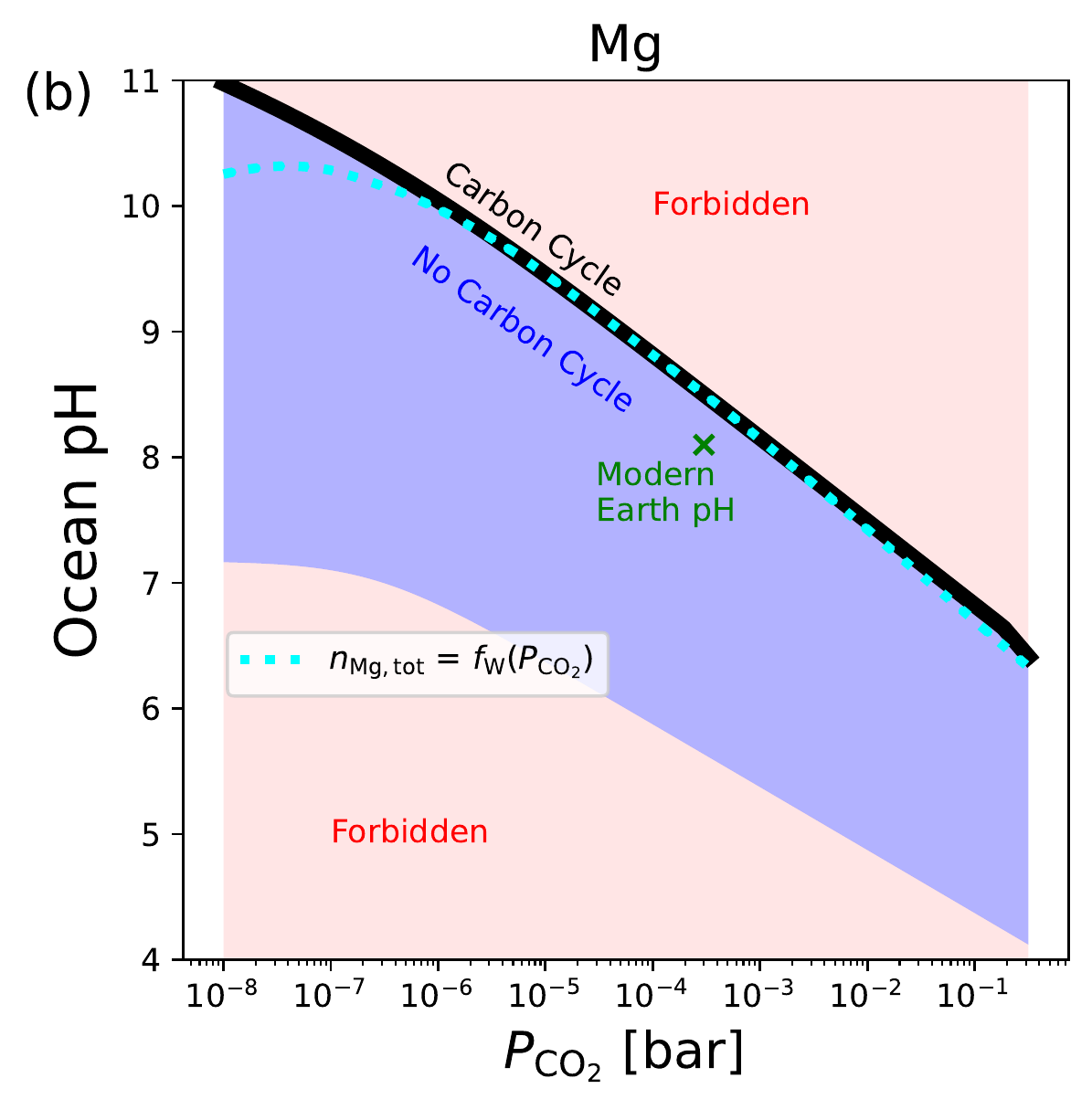}
\includegraphics[width=0.375\textwidth]{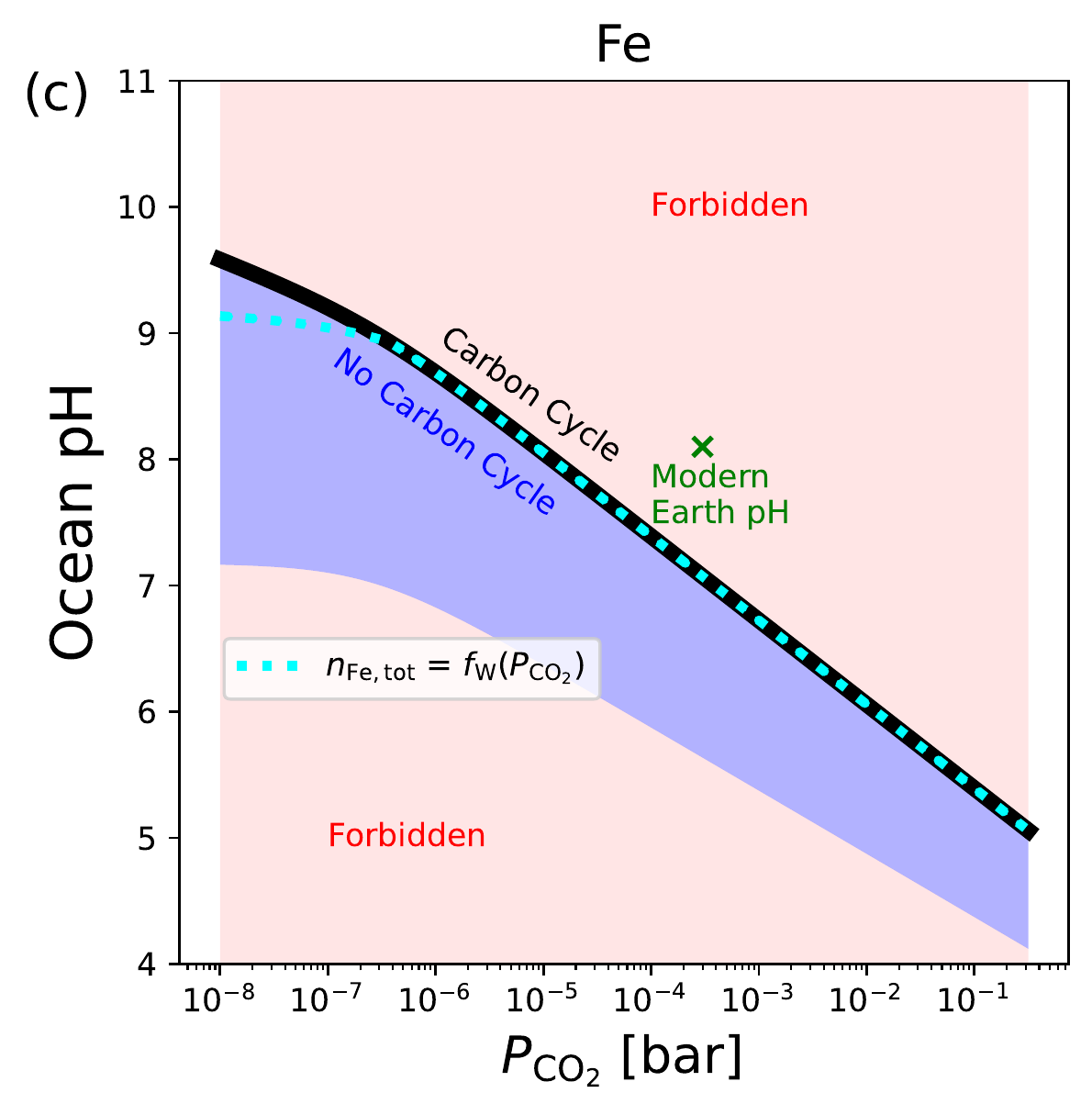}
\caption{ Sensitivity of ocean pH to $P_{\rm CO_2}$  at $T$ = 288~K for pure (a) Ca, (b) Mg, (c) Fe systems. Upper and lower bounds of ocean pH are represented by the blue shaded region. Pink shaded regions are forbidden. }
\label{fig:fig2}
\end{center}
\end{figure}

\section{Results and Discussion} \label{sec:results}

We consider the ocean pH to be determined by the chemical dissolution of atmospheric carbon dioxide in a well-mixed ocean, which occurs at the atmosphere--ocean interface. The chemical dissolution of CO$_2$ is governed by the reaction between water and CO$_2$ to produce H$^+$, HCO$_3^-$ and CO$_3^{2-}$ ions (Sect. \ref{sec:methods}). As $P_{\rm CO_2}$ increases, the ocean becomes more acidic. We consider an atmospheric surface pressure of 1~bar, but allow the atmospheric carbon dioxide content to vary via $P_{\rm CO_2}$. Atmospheric surface pressures up to 100~bar have a negligible effect on our results and those between 100--1000~bar exhibit a small effect (Fig.~\ref{fig:figA2}a). 

\begin{figure}
\begin{center}
\includegraphics[width=0.475\textwidth]{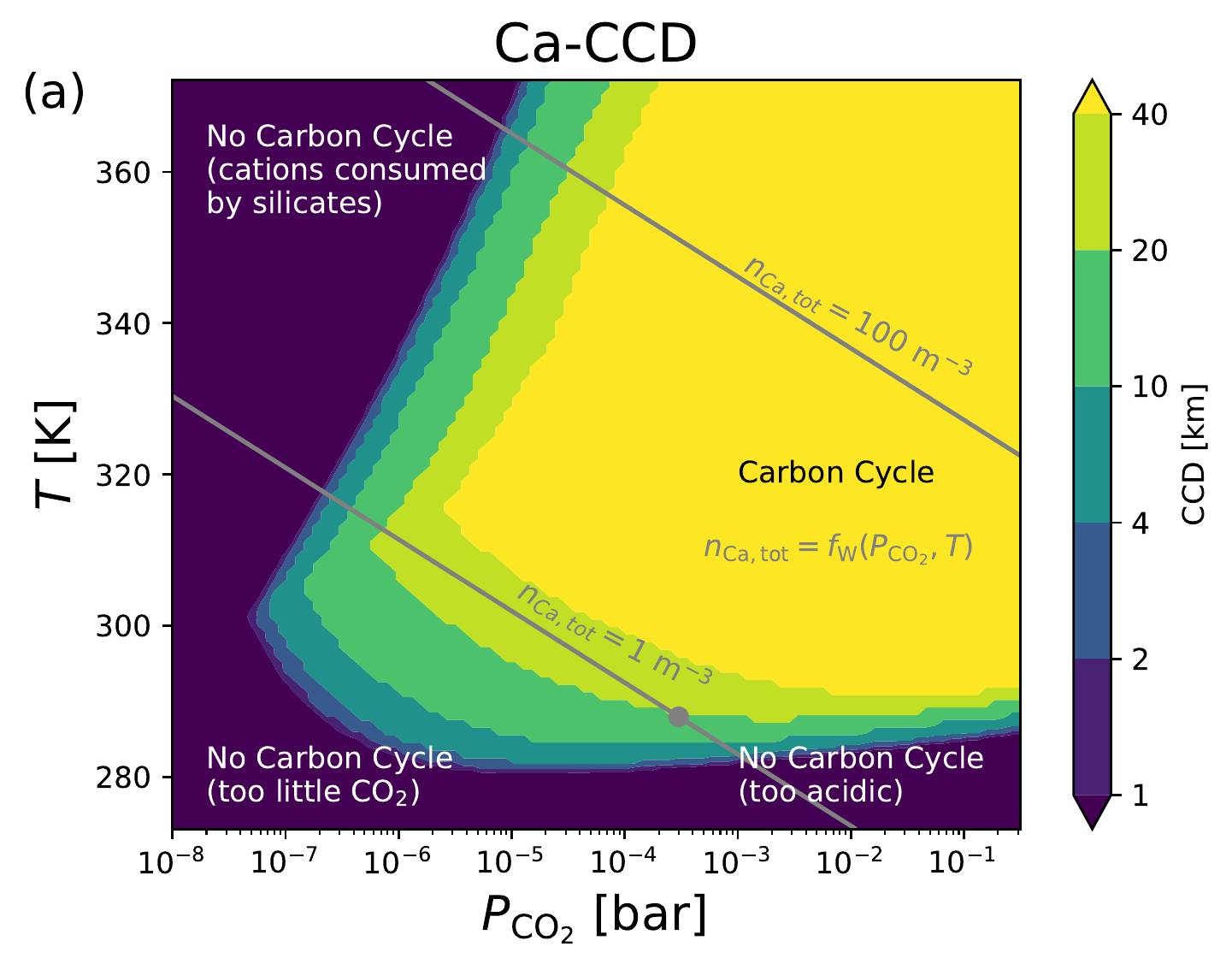}
\includegraphics[width=0.475\textwidth]{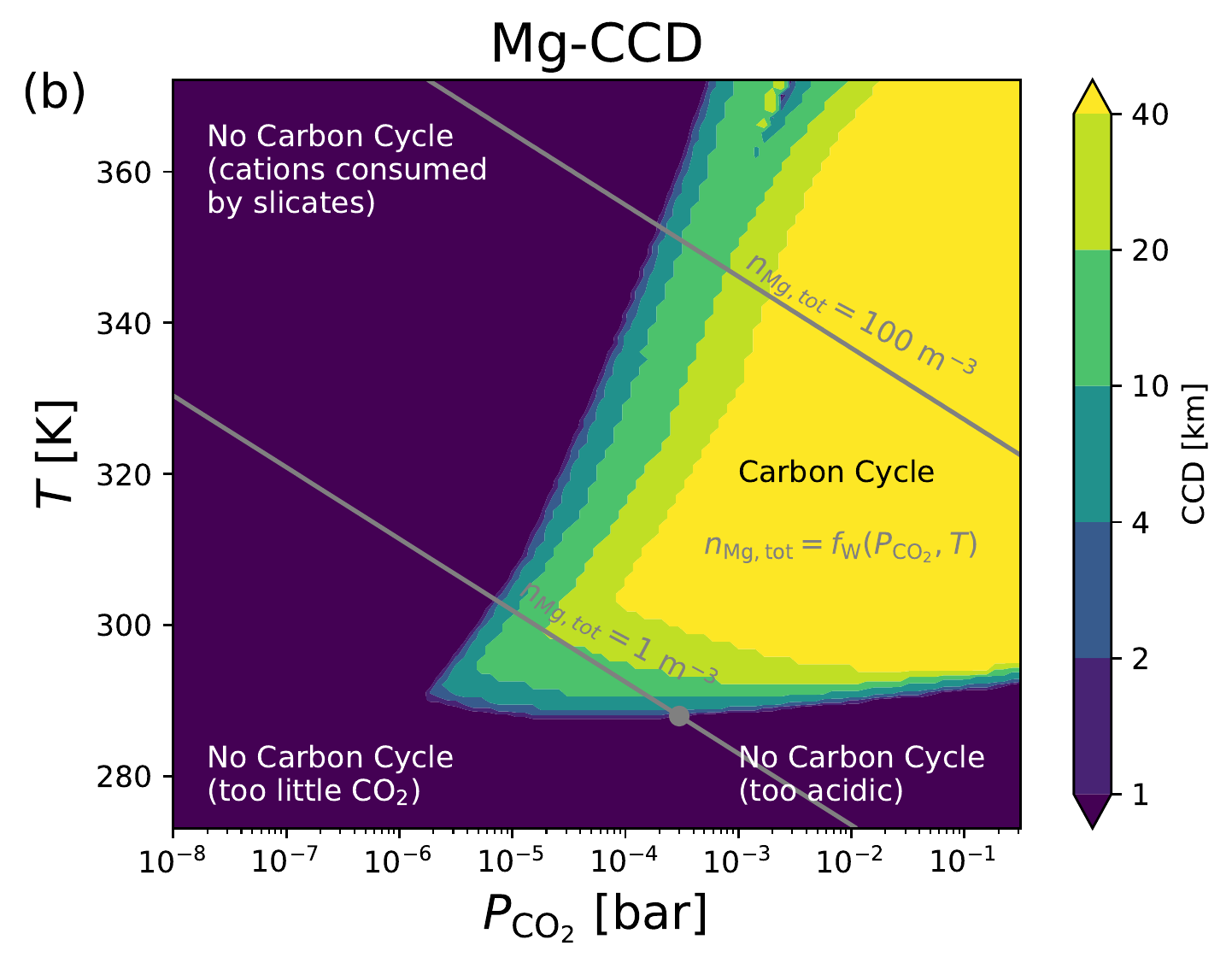}
\includegraphics[width=0.475\textwidth]{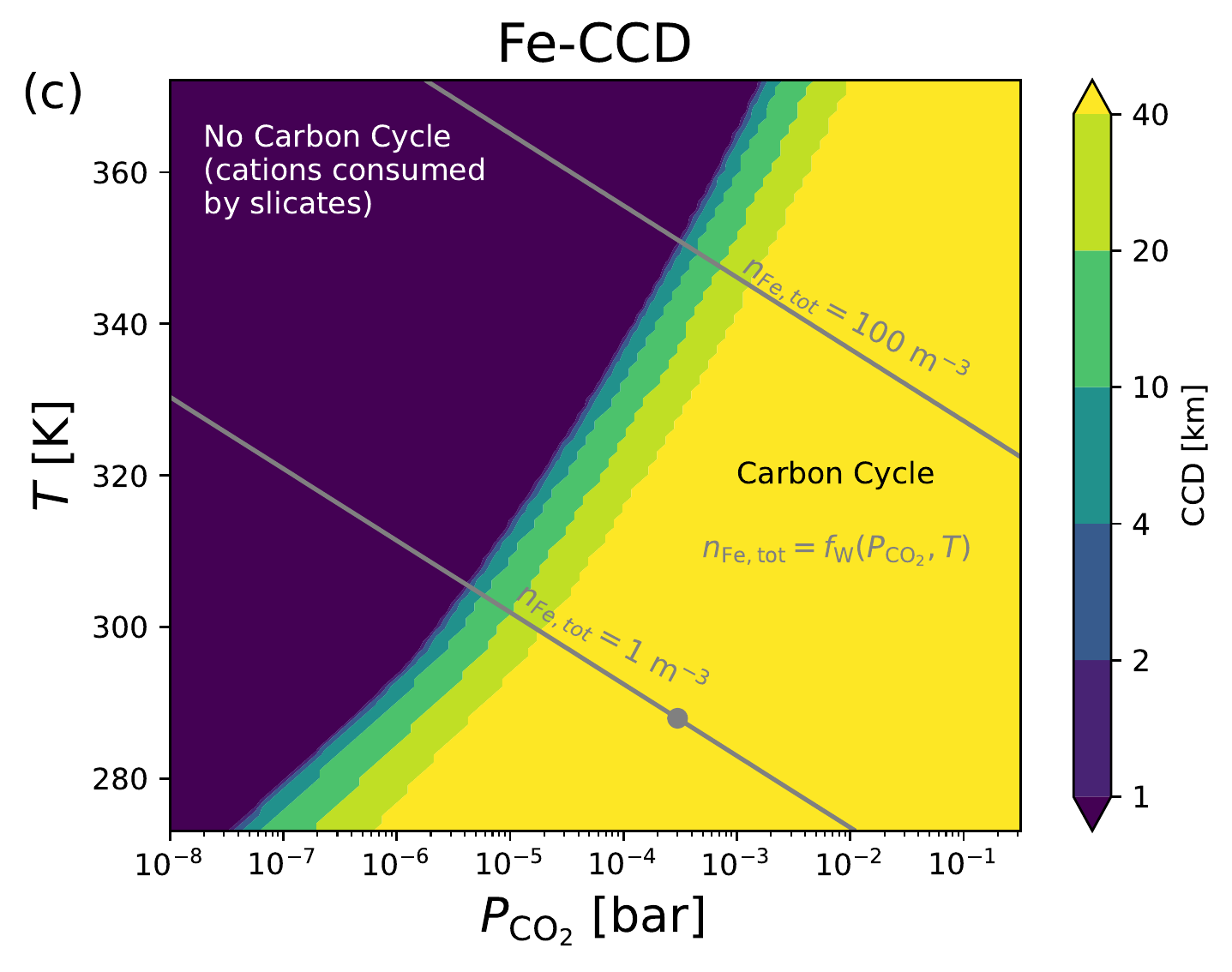}
\caption{Carbonate compensation depth (CCD) as a function of $P_{\rm CO_2}$ and $T$ ($P_{\rm atm} = 1$~bar) for (a) Ca, (b) Mg and (c) Fe systems. Gray contours represent the weathering-dependent cation number density as a function of $P_{\rm CO_2}$ and $T$ (Eq.~\ref{eq:funcCa}). Gray disc denotes modern Earth $P_{\rm CO_2}$ and $T$. }
\label{fig:fig3}
\end{center}
\end{figure}

\begin{figure}
\begin{center}
\includegraphics[width=0.485\textwidth]{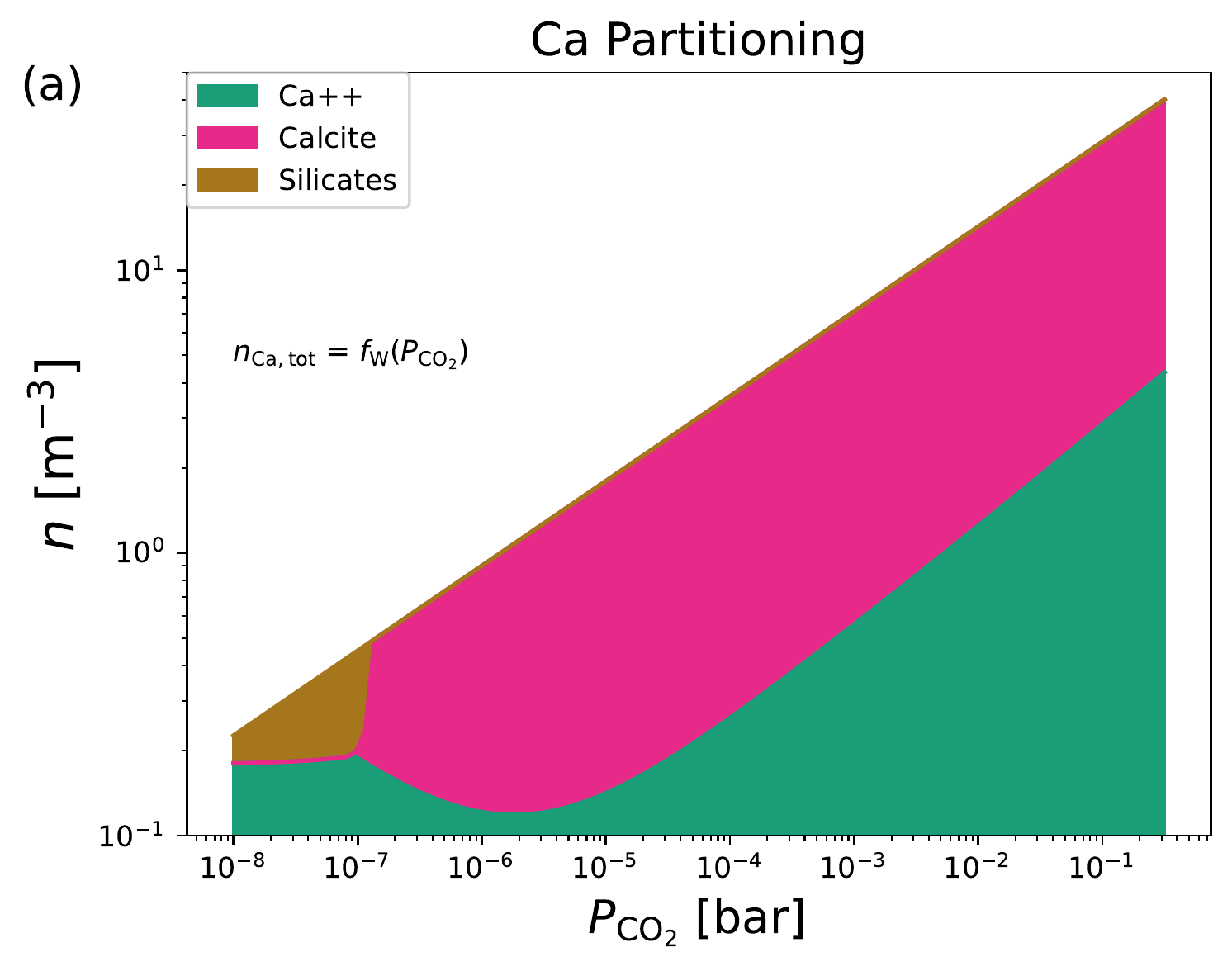}
\includegraphics[width=0.485\textwidth]{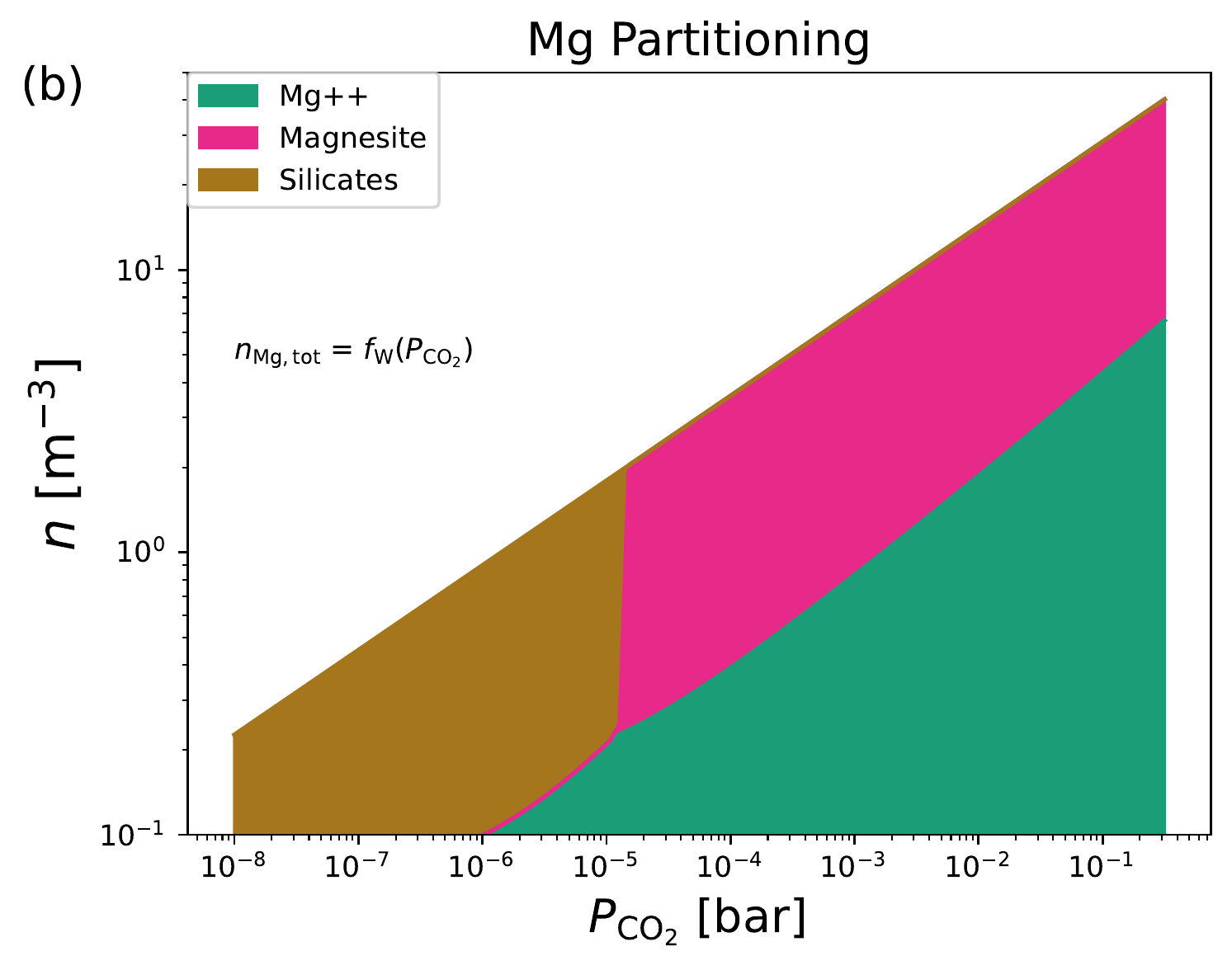}
\includegraphics[width=0.485\textwidth]{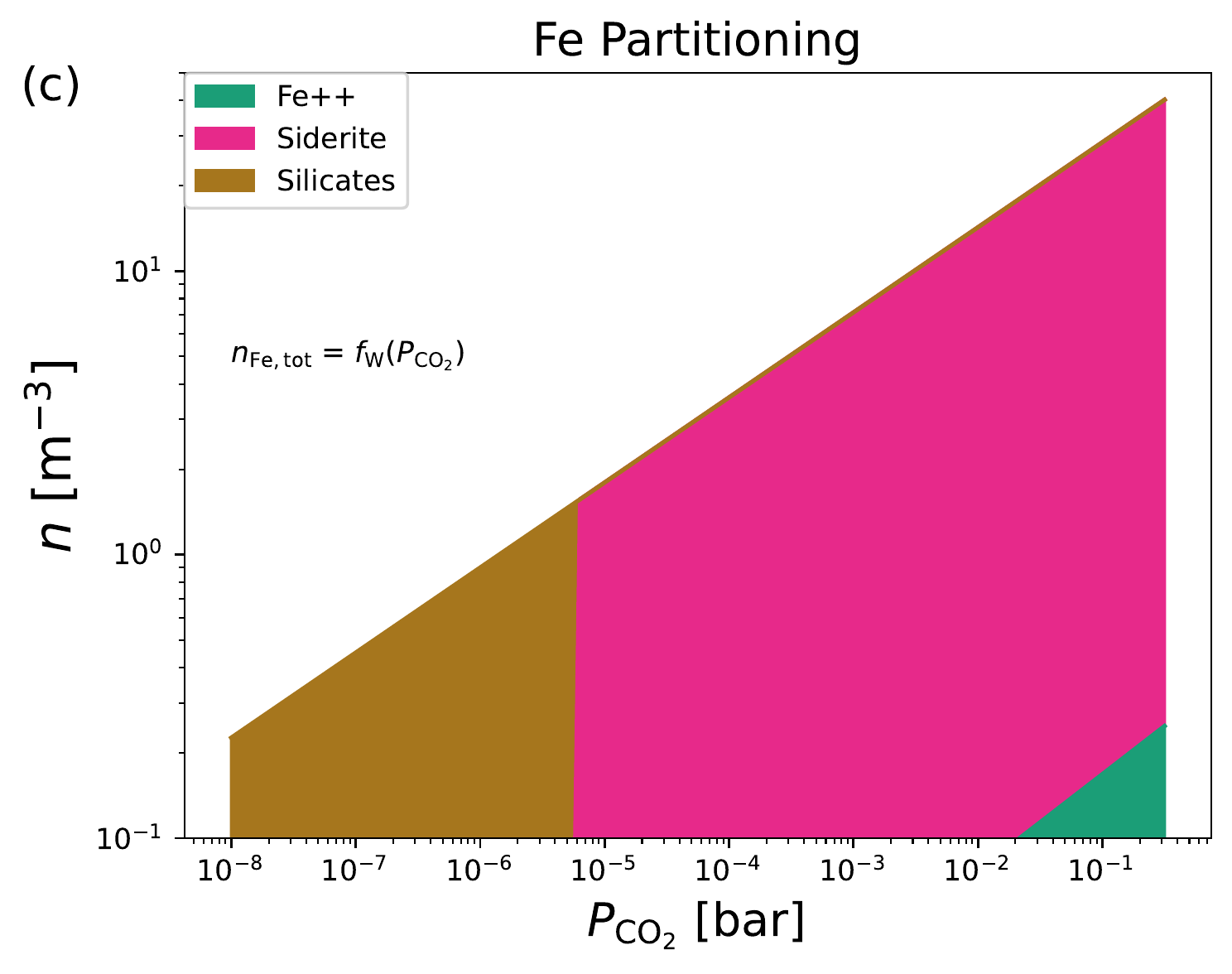}
\caption{Partitioning of (a) Ca, (b) Mg and (c) Fe in aqueous, carbonate and silicate phases as a function of $P_{\rm CO_2}$ at $T = 310$~K ($P_{\rm atm} = P_{\rm oc} = 1$~bar) in pure Ca, Mg and Fe systems, respectively.  }
\label{fig:fig4}
\end{center}
\end{figure}

For a given value of $P_{\rm CO_2}$, the ocean pH is bounded between two limits (Fig.~\ref{fig:fig2}a). The ocean pH is restricted to a narrow range between 7--11 at $P_{\rm CO_2} = 0.01$~$\mu$bar and 4--7 for $P_{\rm CO_2} = 0.1$~bar. These ocean pH ranges are consistent with the inferences for Earth's history, transitioning from an acidic ocean during the Archean at high $P_{\rm CO_2}$ to an alkaline ocean at present-day $P_{\rm CO_2}$  \citep{2017Sci...355.1069H,2018PNAS..115.4105K}. The lower limit corresponds to the complete absence of divalent cations and thus it is independent of the carbonate system under investigation (Sect. \ref{sec:methods}). The upper limit corresponds to the saturation of calcium cations in ocean water such that more weathering does not produce further changes in pH and simply produces more calcite. This upper limit is buffered by the precipitation of carbonates and hence it results in one solution of ocean pH when the carbon cycle is operational for a given carbonate system and $P_{\rm CO_2}$. Both upper and lower limits of ocean pH follow a slope of approximately --0.5 as a function of $P_{\rm CO_2}$ (see Sect. \ref{sec:methodsAnalytic}). Between these two limits, the number density of calcium cations is below the threshold to precipitate carbonates onto the ocean floor; thus, the carbon cycle is not operational.

Due to their high condensation temperatures, the relative abundances of refractory elements observed in the photosphere of stars are expected to be mirrored in the rocky exoplanets they host \citep{2010ApJ...715.1050B,2015A&A...580A..30T}.  For example, the calcium-to-magnesium ratio of the solar photosphere and Earth are 0.062 and 0.066, respectively \citep{2003ApJ...591.1220L,2012Icar..221..859E}. The relative abundances of Ca, Mg and Fe, measured from the spectra of stars, vary by up to an order of magnitude.  For example, Ca/Mg=0.02--0.2 and Ca/Fe=0.04--0.2 in the Hypatia catalogue of more than 7000 stars \citep{2014AJ....148...54H}. Furthermore, carbonates involving Mg and Fe are known to have formed during Earth's history: e.g., magnesite (MgCO$_3$) and siderite (FeCO$_3$); these carbonates have dissolution properties that differ from those of calcite. Siderite could have played a key role in locking up CO$_2$ in carbonates on Earth during the Archean \citep{1995Natur.378..603R,2010Sverjensky}. We calculate ocean pH for the pure Mg and Fe systems in addition to the Ca system (Fig.~\ref{fig:fig2}b,c). The upper limit of ocean pH for a given $P_{\rm CO_2}$ varies when considering systems with purely Ca, Mg or Fe as the source of weathering cations. The upper limit of ocean pH for the Mg system is only 0.2 higher than for the Ca system, whereas it is more than unity lower for the Fe system.

For $P_{\rm CO_2} < 10$~$\mu$bar, ocean chemistry and hence the CCD is sensitive to the addition of aqueous silica (SiO$_2$) in the ocean (Fig.~\ref{fig:fig3}). Silica is another product of silicate weathering, which enables the locking up of cations in silicate minerals instead of carbonate minerals \citep{1981JGR....86.9776W,2021PSJ.....2...49H}. For instance, for $T > 300$~K and $P_{\rm CO_2} < 0.1$~$\mu$bar in the Ca system in the presence of aqueous silica, silicates impinge on the stability of calcite (Fig.~\ref{fig:fig4}a) and prevent carbonate precipitation at all depths (Fig.~\ref{fig:fig3}a). In contrast, when no silica is present in the ocean for $T > 300$~K and $P_{\rm CO_2} < 0.1$~$\mu$bar, calcite is stable (Fig.~\ref{fig:figA4}a) and deep CCDs are produced (Fig.~\ref{fig:figA3}a), thereby increasing the parameter-space where the carbon cycle is stable. Similarly, in the Mg and Fe systems, silicates are more stable than carbonates for $P_{\rm CO_2} < 10$~$\mu$bar (Fig.~\ref{fig:fig4}b,c). $P_{\rm CO_2} > 10$~$\mu$bar favours the thermodynamic stability of carbonates over silicates.

Carbon cycle box models of exoplanets often omit self-consistent modelling of ocean chemistry and precipitation of carbonates. Carbonate precipitation is implicitly assumed to persist and is not expected to be a bottleneck for carbon cycling. Our ocean chemistry model can be incorporated directly into carbon cycle box models for exoplanets, which can couple via key parameters, $P_{\rm CO_2}$, $T$, and the carbonate chemistry. Thermochemical equilibrium calculations of our ocean model can be used to determine the carbon fluxes into or out of the near-surface reservoirs. The carbon cycle box models can also be informed of the effect of ocean chemistry and ocean depth on the efficiency of carbon degassing and recycling.

Upcoming observations of terrestrial exoplanets from the James Webb Space Telescope, Atmospheric Remote-sensing Infrared Exoplanet Large-survey and Extremely Large Telescopes will put constraints on their atmospheric composition, for instance, the volume mixing ratio of atmospheric carbon dioxide ($P_{\rm CO_2}/P$). Determining the partial pressure of carbon dioxide ($P_{\rm CO_2}$) requires the atmospheric surface pressure ($P$) which is not easily constrained. Nonetheless, our thermodynamic calculations provide strong constraints on ocean chemistry in the presence or absence of magnesium, calcium or iron carbonates; the relative abundances of these carbonate-forming elements in planetary systems can be deduced from observations of stellar photospheres.  Our results suggest that the carbon cycle will operate robustly on chemically-diverse terrestrial exoplanets exhibiting silicate weathering. This implies that the search for life from exoplanets with temperate climates or biospheres will benefit by broadening the target list to planets that are more chemically diverse than Earth. \\

We thank the anonymous referee for their valuable comments that improved the quality of this paper. We acknowledge financial support from the European Research Council via Consolidator Grant (ERC-2017-CoG-771620-EXOKLEIN, awarded to K. Heng) and the Center for Space and Habitability, University of Bern. We thank Allan Leal for the support with \texttt{Reaktoro}.

%



\section*{Data availability}

All data generated or analysed during this study are included in the published article.

\section*{Code availability}

\texttt{OCRA} (Ocean Chemistry with Reaktoro And beyond): the open-source code developed in this work is hosted at \url{https://github.com/kaustubhhakim/ocra}. \texttt{OCRA} v1.0 was used in this study and is also available on Zenodo \citep{2022ocra}.

\software{\texttt{numpy} \citep{2020numpy}, \texttt{scipy} \citep{2020SciPy-NMeth}, \texttt{pandas} \citep{2020pandas}, \texttt{astropy} \citep{2013A&A...558A..33A,2022ApJ...935..167A}, \texttt{matplotlib} \citep{2007matplotlib}, \texttt{Reaktoro} \citep{2015reaktoro}
          }




\appendix

\twocolumngrid

\renewcommand{\thefigure}{A\arabic{figure}}
\setcounter{figure}{0}

\section{Analytical solution of ocean pH and \textit{P--T} sensitivity} \label{app:A}

The analytical solution for the upper limit of ocean pH is derived from the relations between the equilibrium constants and reactants and products (assuming water activity to be unity in diluted solutions) of reactions described by Equations \ref{eq:Ca++} and \ref{eq:ChemicalCO2B},

\begin{equation}
\label{eq:Ca++eq}
K_{9} = \frac {n_0^2} { n_{\rm{Ca^{2+}}} n_{\rm{CO_3^{2-}}} },
\end{equation}

\begin{equation}
\label{eq:ChemicalCO2Beq}
K_{16} = \frac {n^2_{\rm H^+} n_{\rm{CO_3^{2-}}} } { P_{\rm CO_2} n_0^3 }.
\end{equation}
By eliminating the carbonate ion number density from these two equations, proton number density is
\begin{equation}
\label{eq:upperLimitH+}
\frac{n_{\rm H^+}}{n_0} = \left( P_{\rm CO_2} K_{9} K_{16} \frac { n_{\rm{Ca^{2+}}} } {n_0} \right)^{1/2}
\end{equation}
Because the pH is given by 
\begin{equation}
\label{eq:pHeq}
\mathrm{pH} = - \log (n_{\rm H^+}/n_0),
\end{equation} 
the analytical upper limit of ocean pH is Equation \ref{eq:pHUpper}. 

The analytical solution for the lower limit of ocean pH is derived from the the equilibrium constant of the reaction described by Equation \ref{eq:ChemicalCO2}, 
\begin{equation}
\label{eq:ChemicalCO2eq}
K_{3} = \frac{n_{\rm H^+} n_{\rm HCO_3^-}} {P_{\rm CO_2} n_0^2}.
\end{equation}
Then the proton number density is
\begin{equation}
\label{eq:lowerLimitH+}
\frac{n_{\rm H^+}}{n_0} = \frac{K_{3} P_{\rm CO_2} n_0}{n_{\rm HCO_3^-}}
\end{equation}
Thus, the lower limit of ocean pH is given by Equation \ref{eq:pHLower}.

\begin{figure}
\begin{center}
\includegraphics[width=0.48\textwidth]{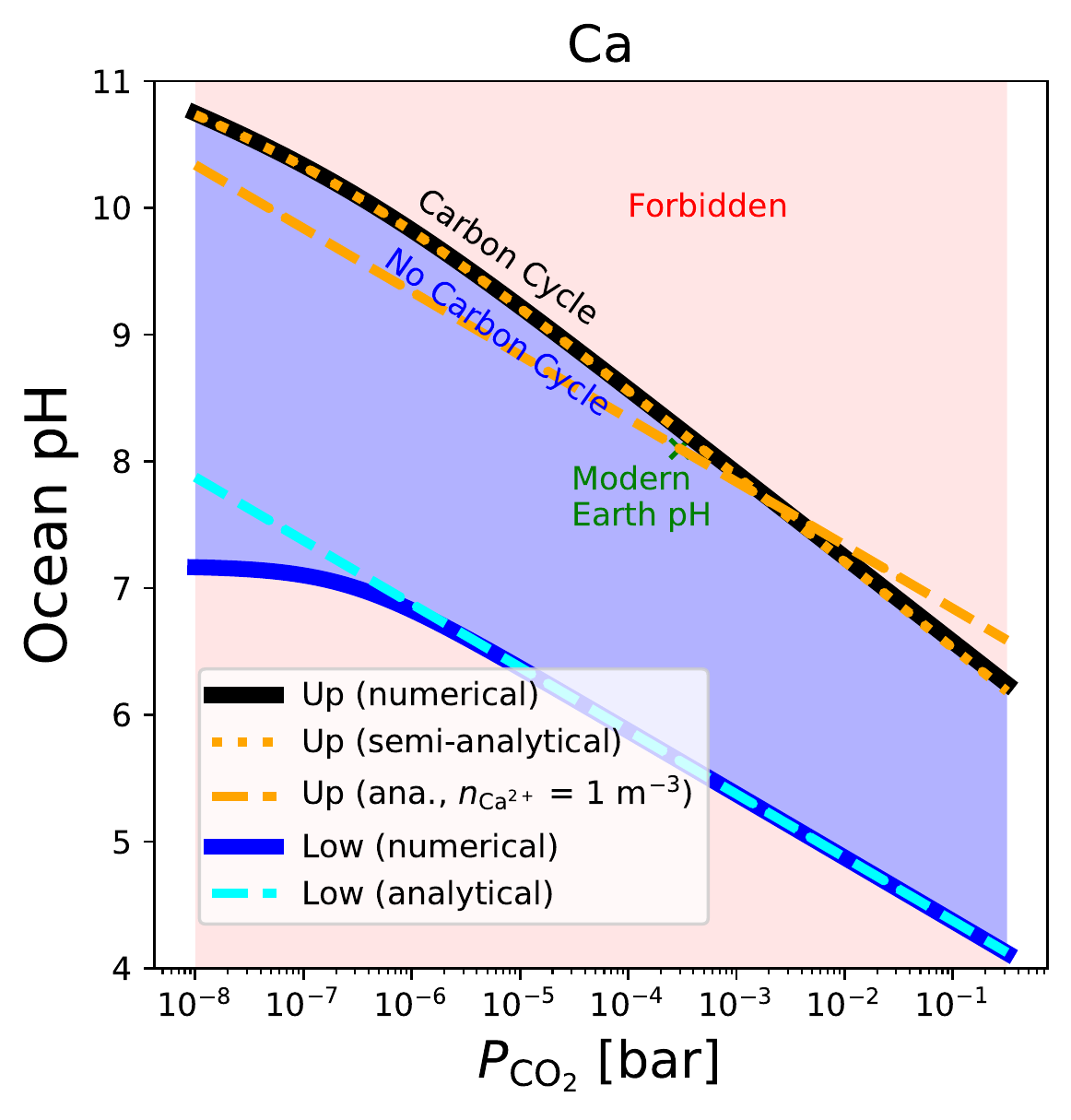}
\caption{Numerical, analytical and semi-analytical solutions of the upper and lower limits of ocean pH in the Ca system.  }
\label{fig:figA1}
\end{center}
\end{figure}

\begin{figure}
\begin{center}
\includegraphics[width=0.5\textwidth]{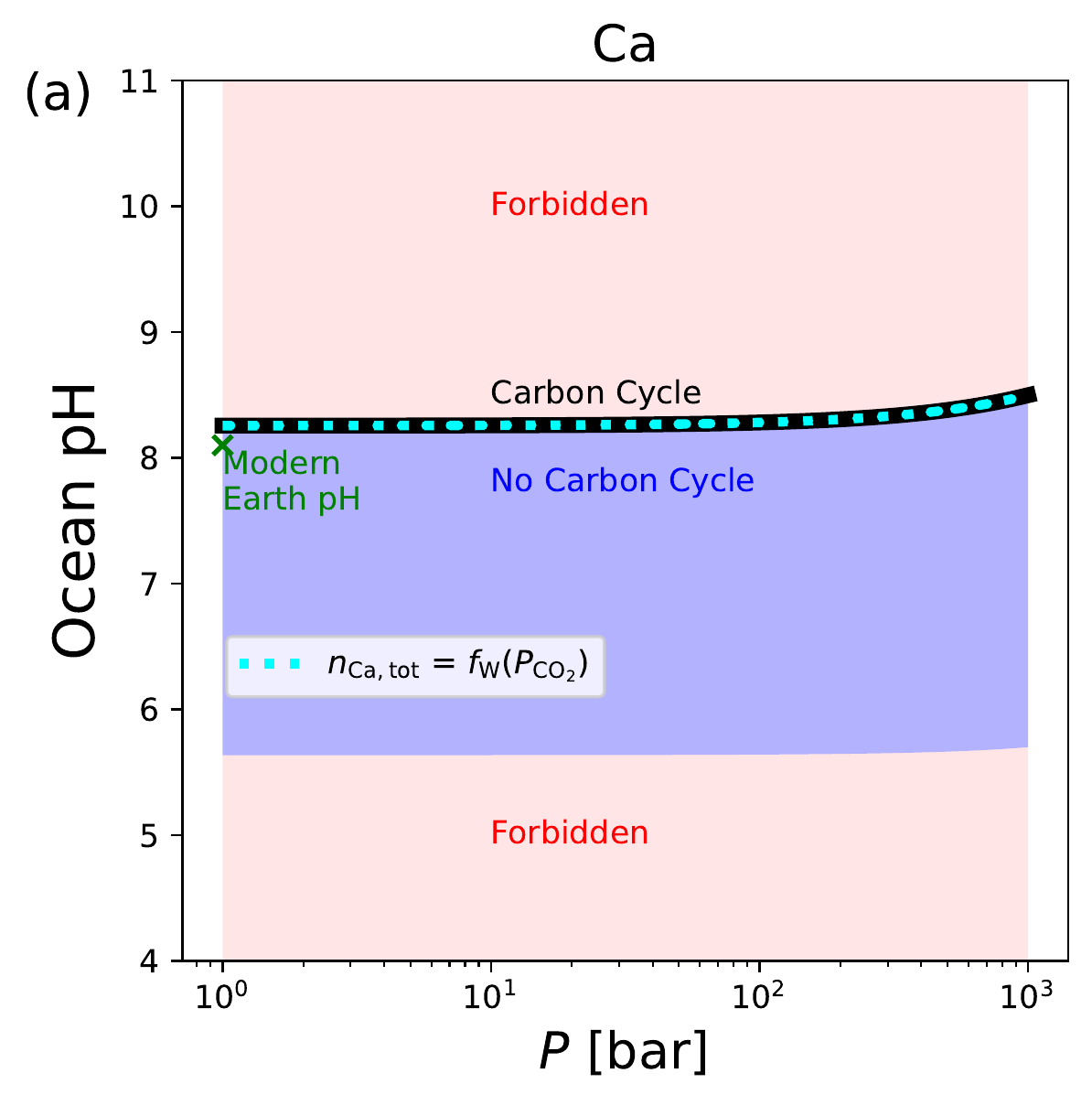}
\includegraphics[width=0.5\textwidth]{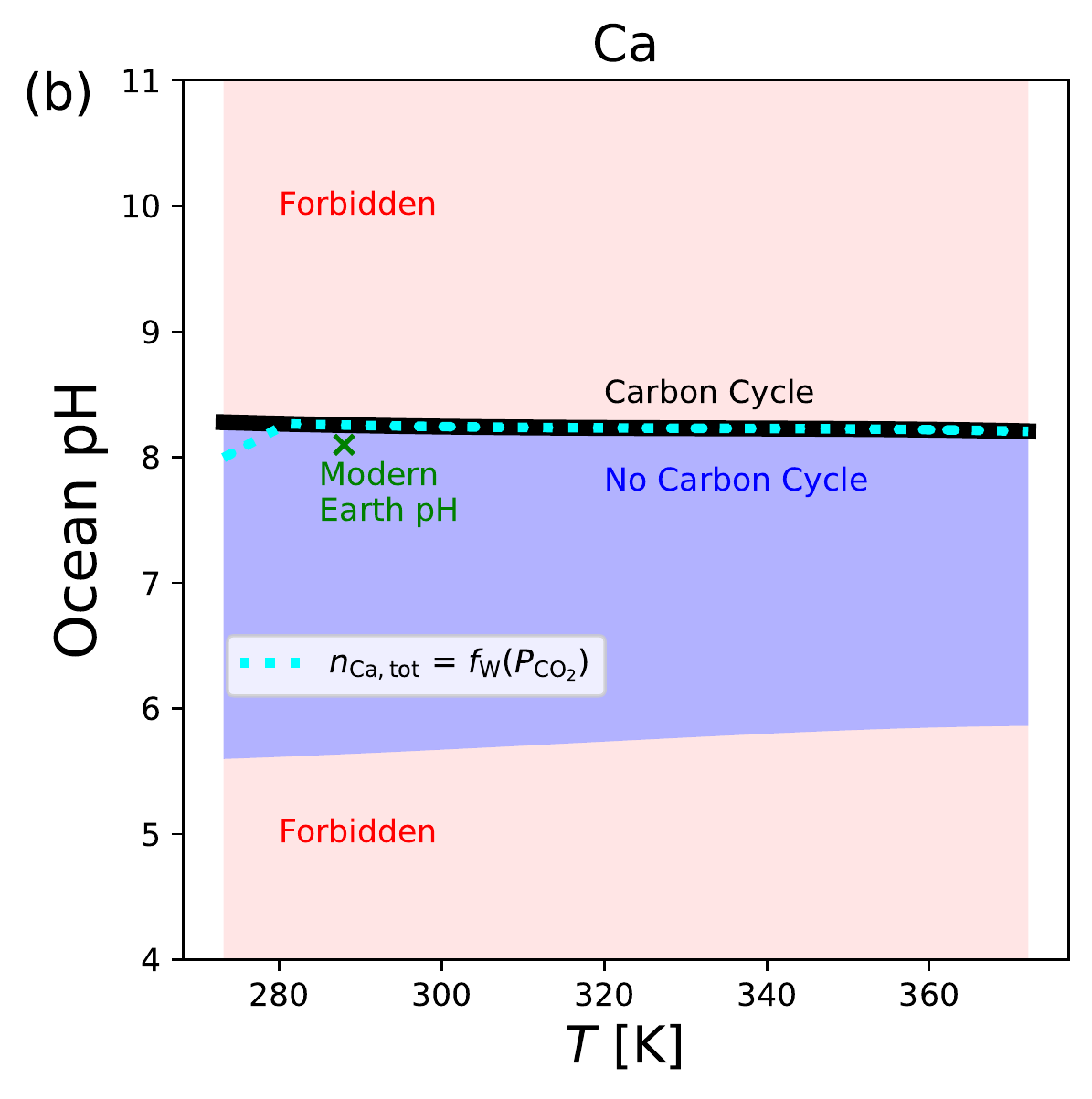}
\caption{The sensitivity of ocean pH to (a) $P$ and (b) $T$ in the Ca-system. }
\label{fig:figA2}
\end{center}
\end{figure}

The analytical solutions of upper and lower limits of ocean pH as a function of $P_{\rm CO_2}$ result in a slope of --0.5 (Fig.~\ref{fig:figA1}). Pressure and temperature have a negligible effect on ocean pH (Fig.~\ref{fig:figA2}).

\newpage

\renewcommand{\thefigure}{B\arabic{figure}}
\setcounter{figure}{0}

\section{CCD without silicate precipitation}

\begin{figure}
\begin{center}
\includegraphics[width=0.495\textwidth]{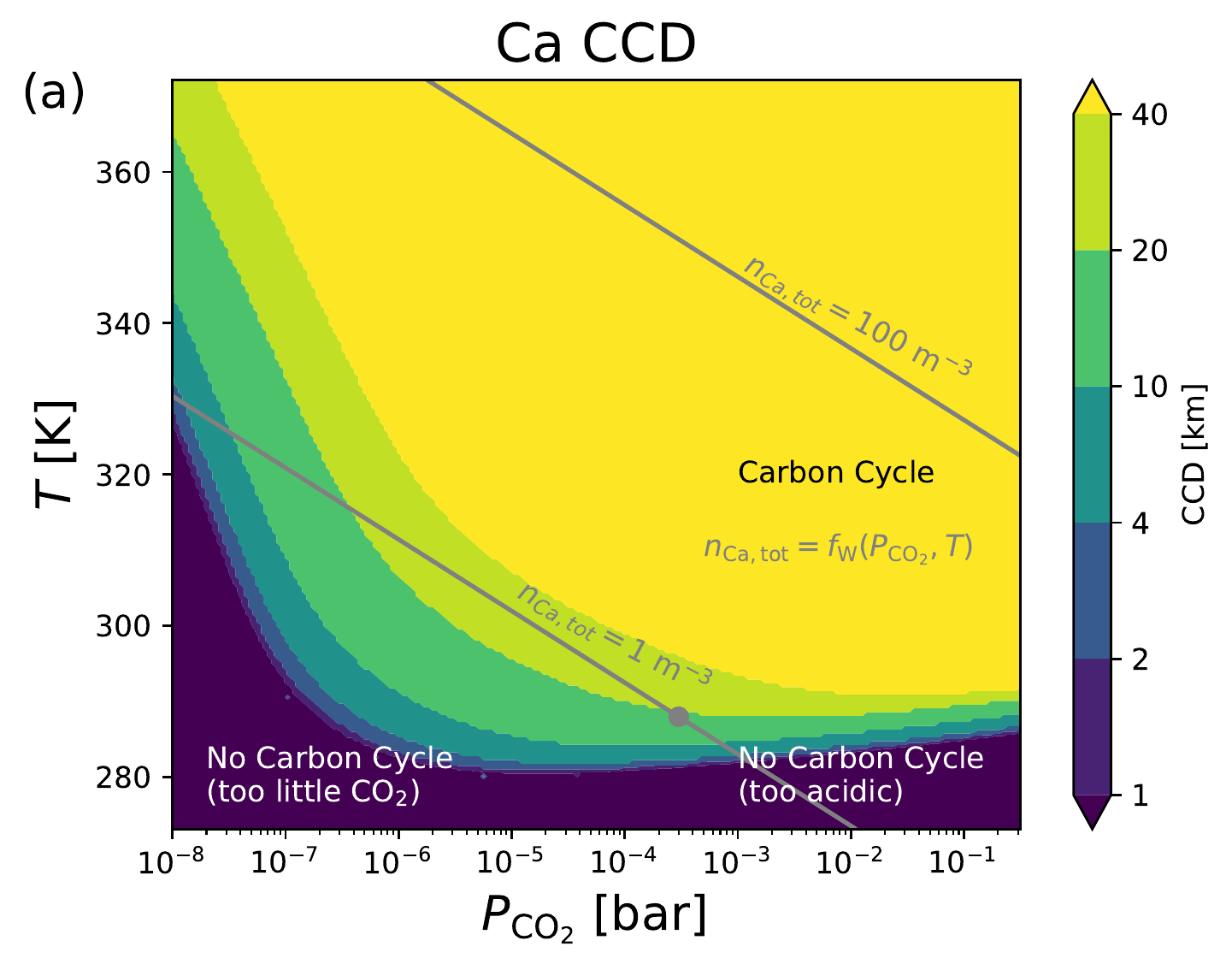}
\includegraphics[width=0.495\textwidth]{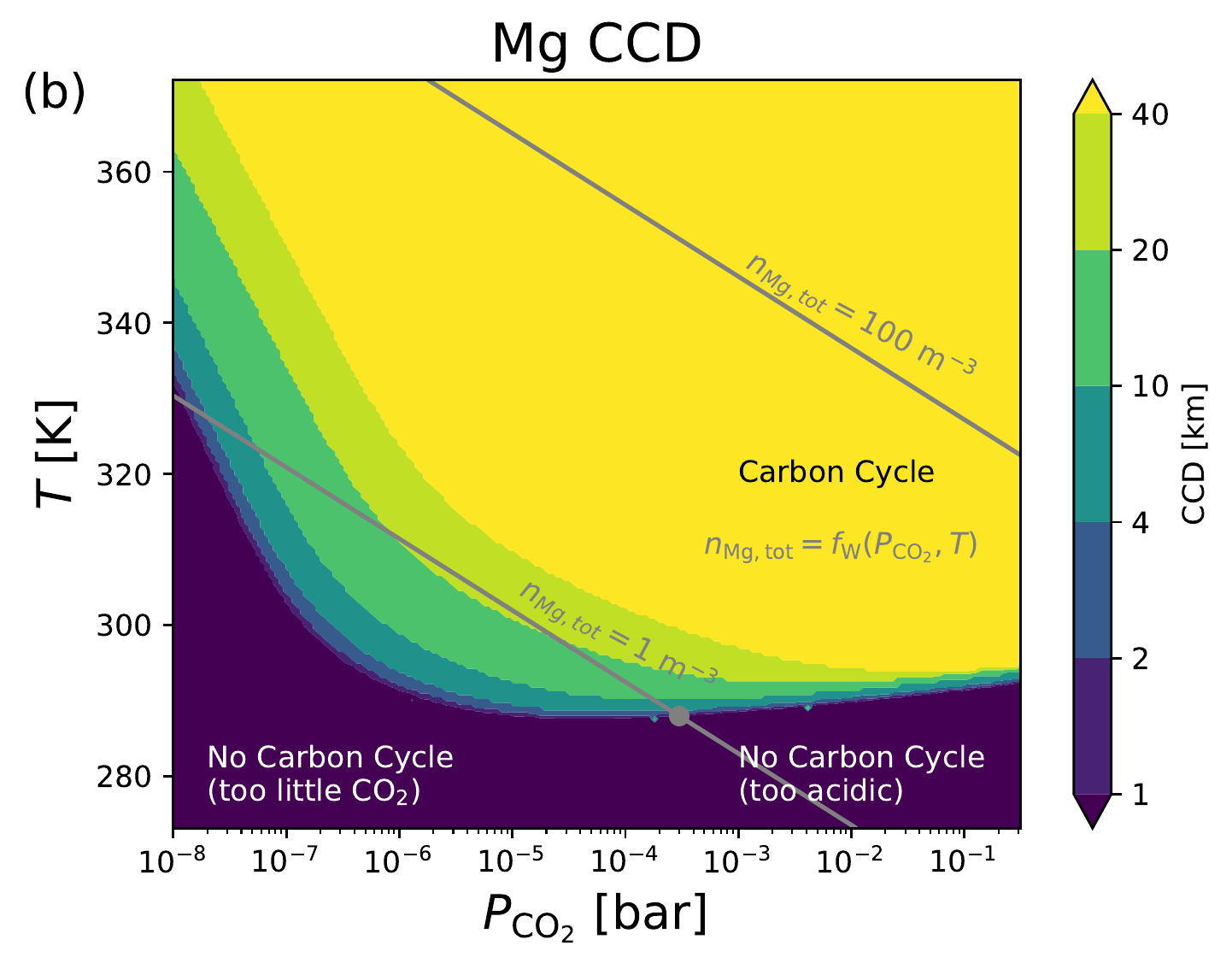}
\includegraphics[width=0.495\textwidth]{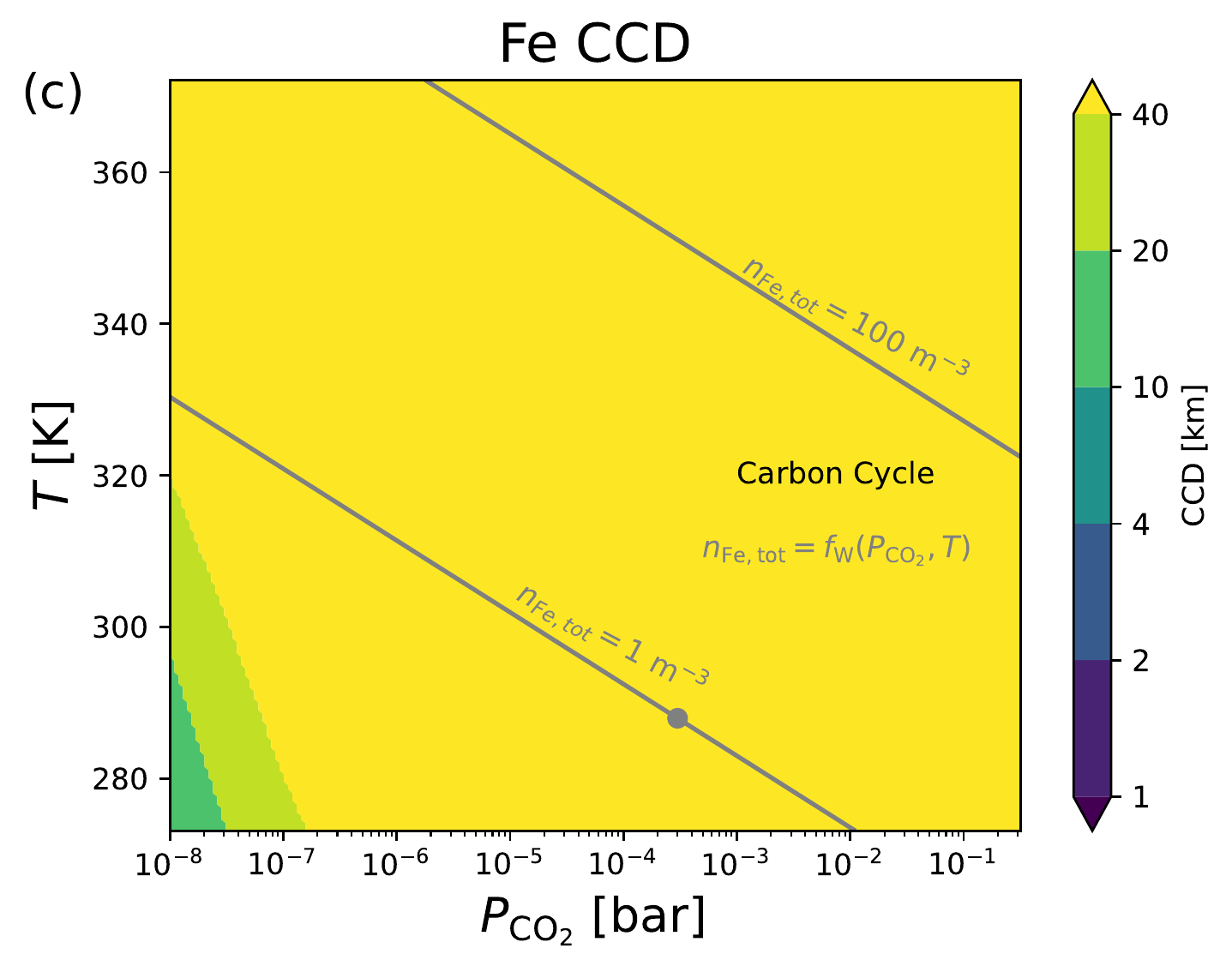}
\caption{Same as Fig.~\ref{fig:fig3} but with no silica $n_{\rm SiO_{2, tot}}$ = 0. }
\label{fig:figA3}
\end{center}
\end{figure}

\begin{figure}
\begin{center}
\includegraphics[width=0.495\textwidth]{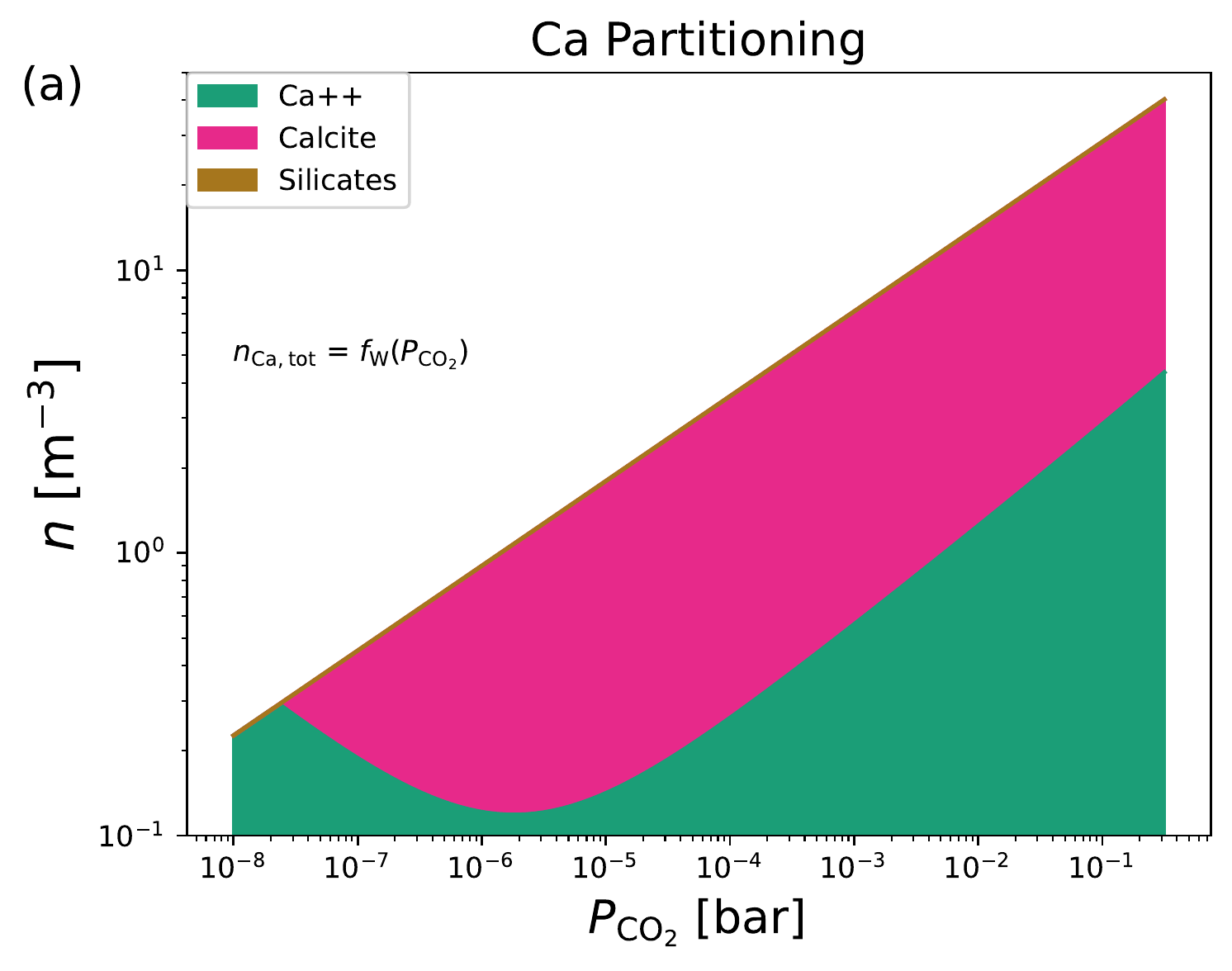}
\includegraphics[width=0.495\textwidth]{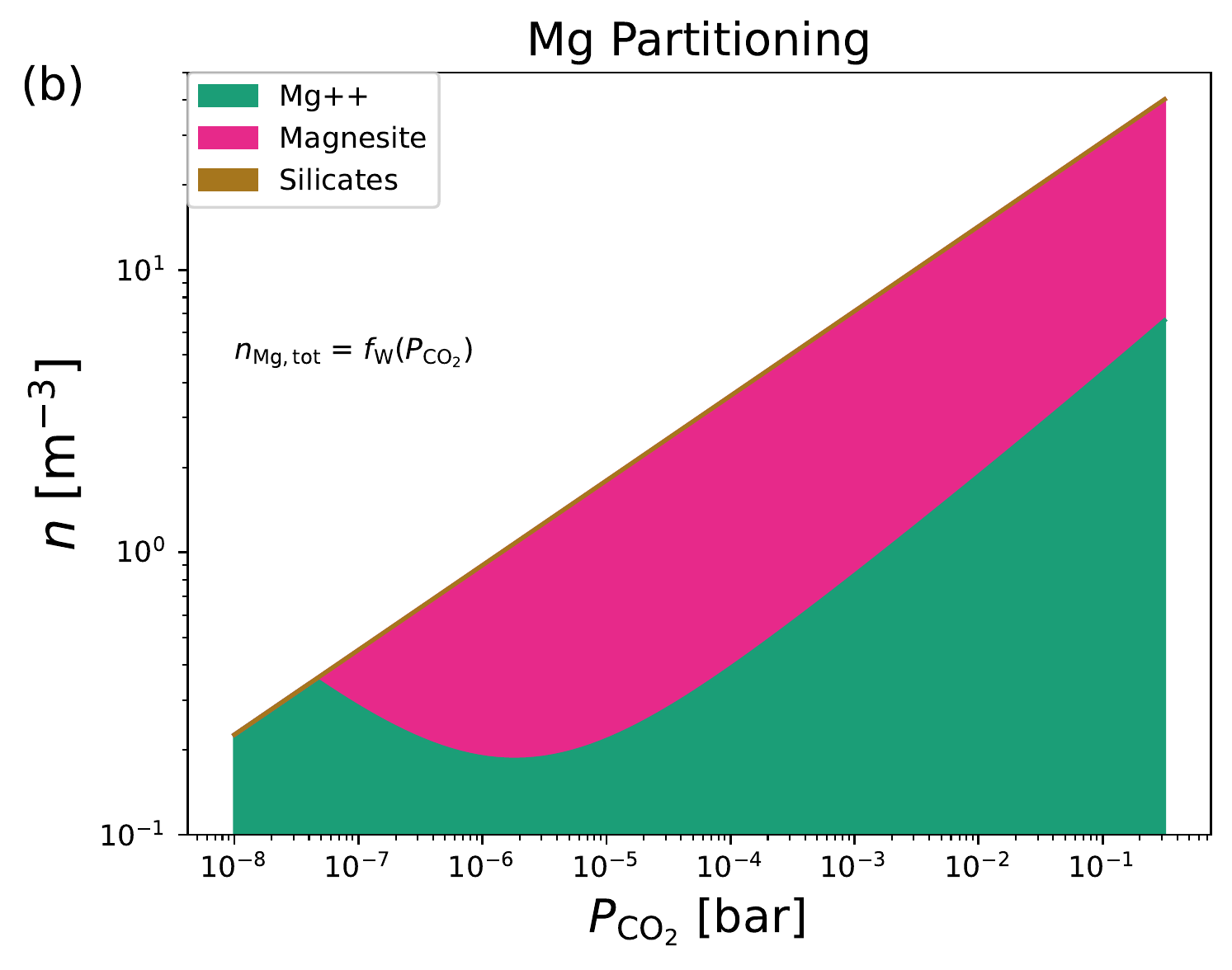}
\includegraphics[width=0.495\textwidth]{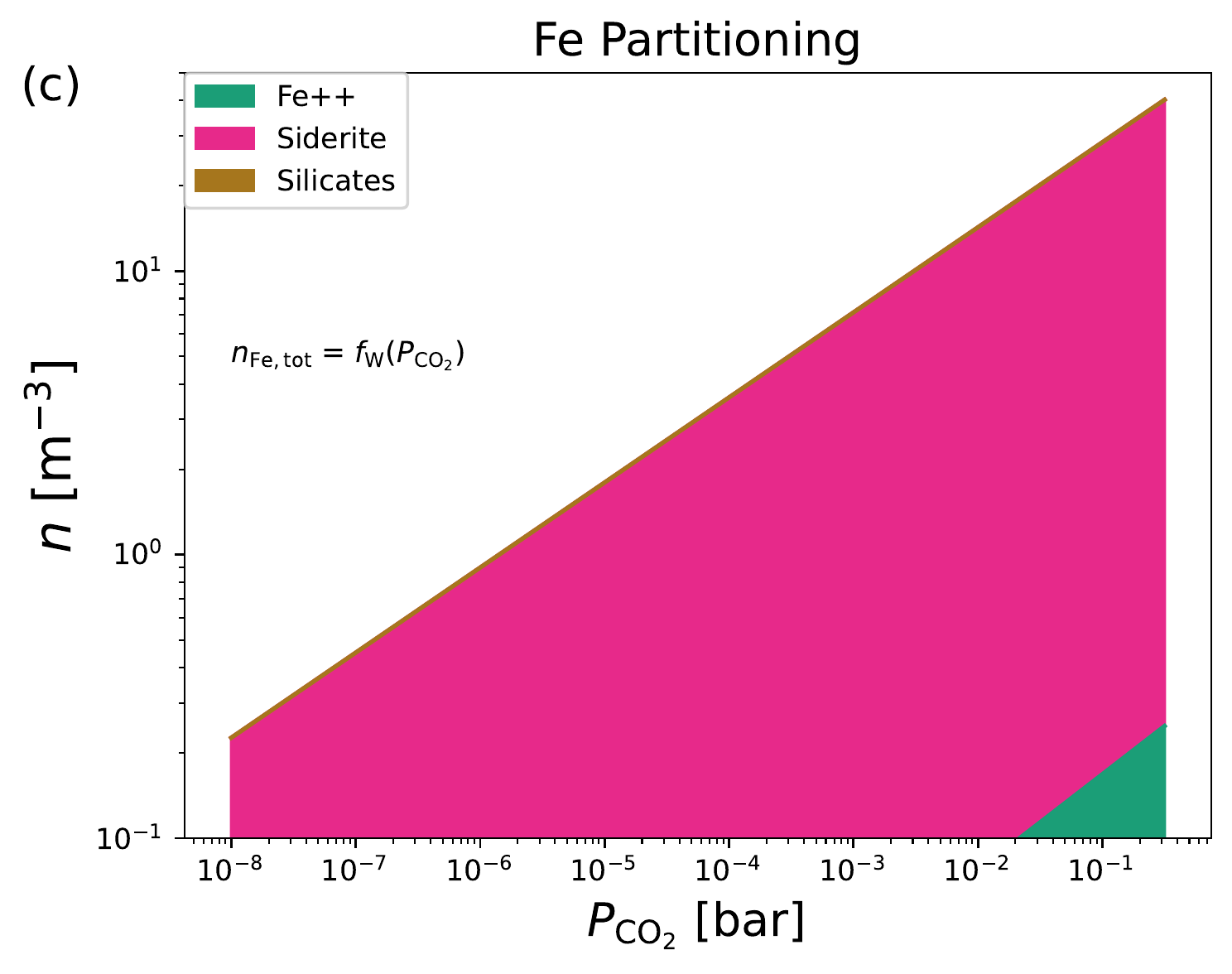}
\caption{Same as Fig.~\ref{fig:fig4} but with no silica $n_{\rm SiO_{2, tot}}$ = 0. }
\label{fig:figA4}
\end{center}
\end{figure}

When no silicates are allowed to precipitate, CCDs for the Ca, Mg and Fe systems become deeper for $P_{\rm CO_2} < 1$~$\mu$bar (Fig.~\ref{fig:figA3}).  This is reflected in the phase stability plots in Fig.~\ref{fig:figA4}.

\clearpage 

\bibliography{references}
\bibliographystyle{aasjournal}



\end{document}